\documentclass[prl,showpacs,twocolumn]{revtex4}% Physical Review B
\usepackage{graphicx}% Include figure files
\usepackage{dcolumn}% Align table columns on decimal point
\usepackage{bm}% bold math

\begin{document}
% define commands for international characters
\catcode`\ä = \active \catcode`\ö = \active \catcode`\ü = \active
\catcode`\Ä = \active \catcode`\Ö = \active \catcode`\Ü = \active
\catcode`\ß = \active \catcode`\é = \active \catcode`\è = \active
\catcode`\ë = \active \catcode`\ô = \active \catcode`\ê = \active
\catcode`\ø = \active \catcode`\ò = \active \catcode`\í = \active
\defä{\"a} \defö{\"o} \defü{\"u} \defÄ{\"A} \defÖ{\"O} \defÜ{\"U} \defß{\ss} \defé{\'{e}}
\defè{\`{e}} \defë{\"{e}} \defô{\^{o}} \defê{\^{e}} \defø{\o} \defò{\`{o}} \defí{\'{i}}

\preprint{APS/123-QED}

\title{An Optically Plugged Quadrupole Trap for Bose-Einstein Condensates}% Force line breaks with \\

\author{D. S. Naik and C. Raman}
 \email{craman@gatech.edu}
%\author{TBD}
%\author{TBD}%
\affiliation{%
School of Physics, Georgia Institute of Technology, Atlanta,
Georgia 30332
}%
\date{\today}% It is always \today, today,
\begin{abstract}
We created sodium Bose-Einstein condensates in an optically
plugged quadrupole magnetic trap (OPT).  A focused, 532nm laser
beam repelled atoms from the coil center where Majorana loss is
significant.  We produced condensates of up to $3 \times 10^7$
atoms, a factor of 60 improvement over previous work
\protect{\cite{davi95bec}}, a number comparable to the best
all-magnetic traps, and transferred up to $9 \times 10^6$ atoms
into a purely optical trap. Due to the tight axial confinement and
azimuthal symmetry of the quadrupole coils, the OPT shows promise
for creating Bose-Einstein condensates in a ring geometry.

\end{abstract}

\maketitle                                %display desired
\vspace{-2 in} Developing new experimental methods has been key to
the exploration of quantum gases. In recent years, successes at
loading atoms directly onto fabricated structures
\cite{hans01,ott01}, evaporation in purely optical traps
\cite{barr01,gran02,webe03}, and transport of atoms from one
vacuum chamber to another \cite{grei01transport,gust02,lewa02}
have created Bose-Einstein condensates (and in some cases
degenerate gases of new species) in environments where enhanced
optical access and proximity to surfaces allows for the discovery
of new phenomena.

Large volume magnetic traps \cite{cornkett99var} are a workhorse
of the field due to their significant capture range compared with
optical traps.  However, they possess an intrinsic trade-off
between tight confinement and optical access, the former requiring
a number of opaque coils to be placed in close proximity to the
atoms. In this work we created Bose-Einstein condensates (BECs)
using an ``optically plugged'' quadrupole magnetic trap (OPT) that
significantly alleviates this trade-off.  An earlier version of
the OPT was used by Ketterle's group to demonstrate the first BEC
in sodium \cite{davi95bec}.  The central result of this paper is a
60-fold improvement in atom number, achieving the 3rd largest
alkali BEC reported \cite{abos01latt,gorl03}. Moreover, although
the OPT does not have a simple harmonic potential energy surface
(see below), we have transferred up to 9 million atoms into a
purely optical trap created by a single, focused infrared laser
beam. This demonstrates that the OPT is an excellent starting
point for BEC experiments.

\begin{figure}[hb]
\begin{center}
\vspace{-0.3 in}
\includegraphics[width = 0.48 \textwidth]{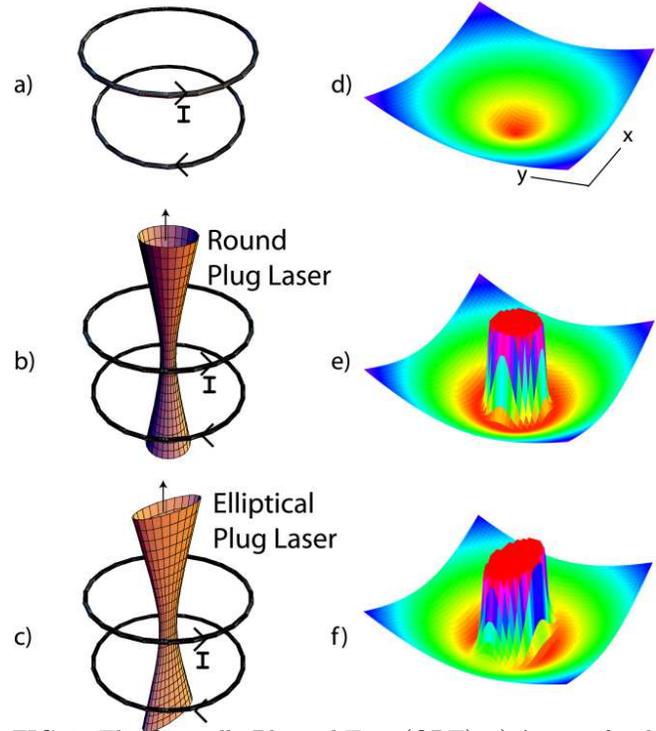}
\end{center}
\vspace{-0.75 in} \caption{The Optically Plugged Trap (OPT). a) A
pair of coils with opposite currents constitutes a quadrupole
trap, which has a hole in the center where atoms are lost by
Majorana spin-flips. b) An optically plugged trap (OPT) is
realized by focusing an intense, blue-detuned laser beam to the
center that repels atoms from the hole, resulting in a
Bose-Einstein condensate free to move along a ring. c) Using an
elliptically shaped focus, the symmetry of the ring is broken, and
two distinct minima form along the minor axis of the ellipse.
About each minimum the atoms experience 3-dimensional, harmonic
confinement with 3 distinct frequencies. (d-f) Potential energy
versus position in the x-y plane corresponding to the trapping
geometries of (a-c).} \label{fig:opt}
\end{figure}

Our OPT should considerably simplify the process of attaining a
BEC, as it requires, apart from the ``plug'' laser, just one pair
of electromagnets driven by a single power supply (this also makes
the magneto-optical trap, or MOT, ensuring exact overlap between
their respective centers).  We can achieve the high loading
efficiency of a magnetic trap without the complexity and
restricted optical access of a multi-coil arrangement typical of
Ioffe-Pritchard (IP) traps such as the cloverleaf design
\cite{mewe96bec}. Moreover, the focusing of an additional, intense
laser beam adds only a minor complexity to the apparatus,
comparable to that required for optical confinement of BECs
\cite{stam98odt}. Our design uses a stable, solid state ``plug''
laser at 532 nm requiring little or no adjustment for several
weeks of operation.  Finally, we show here that it may be possible
to scale down the required laser power to less than 1 watt.

A simple configuration of coils which will trap low magnetic field
seeking particles is the quadrupole trap, formed by a pair of
coils running current in the opposite direction in the so-called
`` anti-Helmholtz'' configuration (see Fig. 1a). Unfortunately, by
itself this trap is not so useful for evaporative cooling--within
a region of $1-2 \mu$m radius near the magnetic field zero at the
trap center, the atoms can spontaneously undergo spin flips and
are lost from the trap \cite{petr95}.  This Majorana loss can be
eliminated if one removes the field zero from the cloud, for
example, using a fast, rotating bias field \cite{petr95,ande95},
or alternately, using a Ioffe-Pritchard design which has a finite
bias field at the trap minimum \cite{mewe96bec}.  Our approach is
based on an idea of Ketterle to use the optical dipole force of a
blue-detuned laser beam to repel atoms from the region containing
the hole \cite{davi95bec}. The resulting potential energy surface
depends on both laser and magnetic fields, and the minimum is
displaced from the coil center so that the atoms experience a
non-zero magnetic field.

Our experimental sequence starts with a Zeeman slowed $^{23}$Na
atomic beam based on a ``spin-flip'' design whose flux is about
10$^{11}$atoms/s.  About $10^{10}$ atoms are loaded in 3 seconds
into a dark MOT in the F=1 hyperfine level \cite{kett93spot}.
Roughly 1/3 of the atoms (the weak-field seekers) are transferred
into the OPT (the magnet and laser beam are turned on
simultaneously), whose axis of symmetry is oriented vertically.
Each coil has 24 windings of 1/8" square cross-section copper
tubing. The average diameter of each coil is 4 inches and their
spacing is 2.25 inches.  A current of 350 A flows through the tube
walls, while cooling water flows through the tube itself, and the
total voltage drop including a high current switch is 20 Volts.
The predicted field gradient is 320 Gauss/cm at this current.
Following the loading of the trap, rf evaporative cooling for 42
seconds resulted in an almost pure Bose-Einstein condensate of
$10-30 \times 10^6$ atoms. In order to achieve such high atom
numbers, we reduced the trap current by a factor of 14 toward the
end of the evaporation stage, thus lowering inelastic losses
associated with high atomic density.  To ensure the magnetic field
zero did not move with respect to the stationary plug beam, it was
imperative to carefully cancel stray magnetic fields.  By
observing the motion of the cloud center and adjusting 3 pairs of
Helmholtz coils, we reduced stray fields to $\lesssim 20$
milliGauss, resulting in $\lesssim 10 \mu$m motion of the field
zero, well below the plug beam diameter.

The plug beam is derived from a 5 Watt, intra-cavity-doubled
Nd:YVO$_4$ laser [Coherent Verdi-5W] at 532 nm.  The laser output
passes through a 40 MHz acousto-optic modulator (AOM)
[Intra-Action Corp.] used for rapid switch-off of the laser beam.
After expanding the beam size by a factor of 3, we focused it into
the vacuum chamber to a beam waist of $\simeq 40 \mu$m using a 500
mm lens.  The green light propagates along the axis of the
quadrupole field (in our notation, the $z$-axis) through the
vacuum chamber and its focus is subsequently imaged onto a CCD
camera with a magnification of 2.5.  We use the same imaging path
to record the atomic absorption using a resonant probe laser at
589 nm.  To suppress the powerful green laser beam during the
absorption measurement, we placed a dichroic mirror, a 589 nm
interference filter and up to 2 bandpass edge filters in the beam
path, for a total suppression of up to 13 orders of magnitude.
%The transmission of the probe light is $50 \%$[??] through the same optics.
The focus of the plug beam was aligned to the center of the atom
cloud in the trap by steering the beam output before the final
focusing lens using a mirror mount with a micrometer actuator.
After coarsely positioning the focus at the center of the circular
absorption image, we performed a fine-tuning by searching for an
enhancement of evaporative cooling, as detailed below.

{\em Ring Trap}.  To understand the potential energy surface of
the combined optical and magnetic fields, we consider an ideal
case where the laser beam has a Gaussian profile with perfect
azimuthal symmetry.  Since the quadrupole potential also possesses
this symmetry (see Fig.1b), this results in a ring-shaped
Bose-Einstein condensate. Neglecting the variation of the laser
beam along $z$ due to the long Rayleigh range of a few
millimeters, the minimum of the combined laser and magnetic
potential is a circle of radius $r_0$ in the $x-y$ plane, which is
located mid-way between the coils (see Fig. 1e).  $r_0$ satisfies
$\frac{\partial}{\partial r}U_0 e^{-2 r^2/W^2}|_{r=r_0} = \mu
B'/2$, with $U_0$ the peak AC stark shift from the plug beam, $W$
the beam waist, $B'$ the axial quadrupole magnetic field gradient,
and $\mu = 1/2 \times$ the Bohr magneton is the magnetic moment
for atoms in the $|F=1,m_F=-1>$ hyperfine state. Typically, $r_0
> W$. Near $r=r_0$ and $z=0$, the potential $U$ varies
harmonically in the $r$ and $z$ directions, with radial curvature
$\partial^2 U/\partial r^2  = \frac{\mu B'}{2 r_0}(4 r_0^2/W^2-1)
$. Along the $z$-direction $\partial^2 U/\partial z^2 = 2 \mu
B'/r_0$.

{\em Harmonic Trap}.  A real laser focus does not have a perfect
Gaussian intensity profile.  The deviation is most significant in
the wings, and for blue-detuned beams this is where the atoms
primarily reside. Small imperfections in a round focus would
result in one or more minima at random locations.  To exert
control over the potential, we artificially broke the symmetry by
creating an elliptical focus whose aspect ratio was $a$.  We
inserted a slit into the beam path before the final focusing lens
to create a spatial profile with roughly the inverted aspect ratio
$1/a$. The resulting beam focus was elliptical, with its minor
axis orthogonal to the slit direction. The ring symmetry was
broken, resulting in two distinct minima on the minor axis $y$ of
the ellipse at $y=\pm y_0$ (see Fig. 1c,1f). Typically, $W_y = 42
\mu$m and $W_x = a W_y$ with $a = 2.5$, $B' = 23$ Gauss/cm in our
decompressed trap, and $U_0/k_B = 65 \mu$K for $2.7$ Watt of laser
power delivered to the atoms.

Near each minimum the potential is harmonic, with 3 distinct
frequencies.  If one neglects gravity and takes $a \rightarrow
\infty$, one has $\omega_x^2 =\mu B'/(2 M y_0)$, $\omega_y^2 =
\omega_x^2 (4 y_0^2/W^2 -1)$, and $\omega_z^2 = 4 \omega_x^2$. For
our parameters, gravity and the finite value of $a$ introduce
corrections of 10 \% to $\omega_{x,z}$, yielding predicted
frequencies of $\omega_{y,x,z} = 2 \pi \times 215,60$ and $125$
Hz, respectively.  In practice, there can also be an asymmetry
between the two minima which is discussed later.

\begin{figure}[h]
\begin{center}
\includegraphics[width = 0.50 \textwidth]{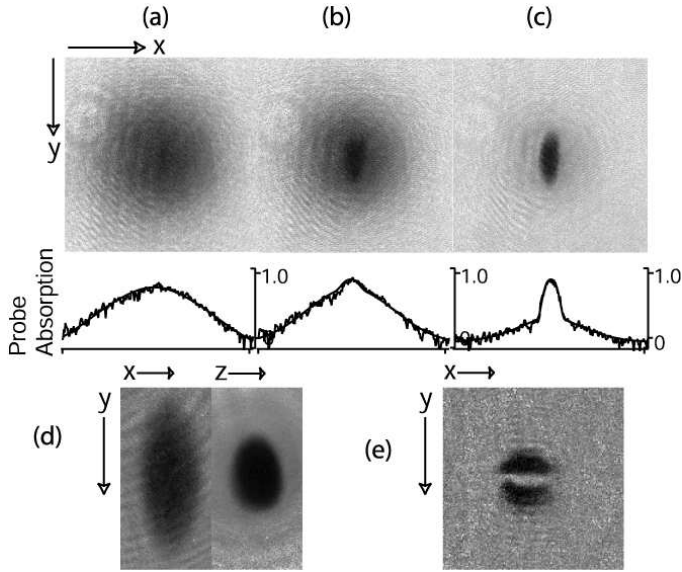}
\end{center}
\vspace{-6mm} \caption{Transition to BEC in the OPT.  Absorption
images taken at final rf frequencies of a) 0.55, b) 0.45, c) 0.30
MHz show the formation of a Bose condensate.  The field of view in
each image is 2.7 $\times$ 2.7 mm.  Below each image is a
horizontal slice through a 2-dimensional bimodal fit to the
absorption data. (d) Views along the $z$ and $x$ directions (40 ms
and 20 ms time-of-flight, respectively) show the 3-dimensional
anisotropic expansion of the condensate. (e) A view in the trap
along the $z$ direction shows two separated clouds near the two
minima of the potential.  Each view in (d) is 1.6 mm (y) $\times$
0.8 mm (x,z), while (e) is 0.8 mm $\times$ 0.8 mm. }
\label{fig:transition}
\end{figure}
\vspace{-3mm} The presence of an elliptical focus dramatically
changed the time-of-flight distribution of the trapped gas.  For
final rf frequencies above $500$ kHz we observed a symmetric
distribution in the $x-y$ plane when imaged along the
$z$-direction, indicating a thermal cloud (Fig. 2a).  Below $500$
kHz we observed a bimodal distribution with an anisotropically
expanding condensate in the shape of a cigar (Figs. 2b,2c).  The
long axis of the cigar was parallel to the minor axis of the
ellipse. While we expect that this axis should rotate if the
ellipse is made to rotate, we could not easily observe this simply
by rotating the slits due to a residual ellipticity in the laser
beam profile.
%When we rotated the orientation of the slits we observed
%that the cigar rotated as well.
At a frequency of $150$ kHz we saw a nearly pure condensate.
Time-of-flight images from the side ($x$-direction) showed that
the condensate expanded anisotropically in 3 dimensions (Fig. 2d).

The measured temperature just above the transition was 1.9 $\mu$K.
The theoretical prediction for a harmonically trapped ideal gas is
$T_c = \frac{\hbar \bar{\omega}}{k_B} (N/1.202)^{1/3}$, where
$\bar{\omega} = (\omega_x \omega_y \omega_z)^{1/3}$
\cite{dalf99rmp}.  For the estimated frequencies above, with $N =
31 \times 10^6$ atoms in 1 (2) wells, we get a temperature of $1.7
\mu$K ($1.3 \mu$K).  The true prediction lies somewhere in
between, since there are two minima whose relative population
depends on the precise laser alignment and profile, which we can
control only with limited precision.  A systematic underestimate
of the atom numbers may be responsible for the higher temperature
measured. The agreement between theory and prediction is
reasonable given the uncertainties in the exact shape of the
potential due to beam alignment as well as the effects of
anharmonicity.

Below the transition, we studied the anistropic expansion of the
condensate.  Assuming harmonic confinement, in the Thomas-Fermi
limit the chemical potential $\mu$ is related to the expanded
cloud sizes through the relation $\mu = \frac{M}{4
t^2}(W_x^2+W_y^2+W_z^2) $ where $W_i$ is the Thomas-Fermi radius
along the $i$-th direction obtained from a parabolic fit to the
time-of-flight image.  For our geometry, roughly 65 \%, 25 \% and
10 \% of the energy is released in the $y$,$z$ and $x$-directions,
respectively.  We estimate $\mu/k_B =$ 600 nK from the measured
cloud sizes.  The Thomas-Fermi prediction is $ \mu =
\frac{15^{2/5}}{2} \left(\frac{N a_s}{\bar{a}}\right )^{2/5} \hbar
\bar{\omega}$ where $N$ is the number of condensed atoms,
$a_s=2.75$nm is the scattering length, and $\bar{a} =
\sqrt{\hbar/M \bar{\omega}}$ is the harmonic oscillator length
\cite{peth02book}.  Estimating the atom number to be $\simeq 20
\times 10^6$ atoms in 1 (2) wells, we compute $\mu/k_B = $ 530 nK
(400 nK).  Both release energy and transition temperature are
slightly higher than the predictions, as would be expected for an
underestimate of the atom number.

\begin{figure}[h]
\begin{center}
\includegraphics[width = 0.50 \textwidth]{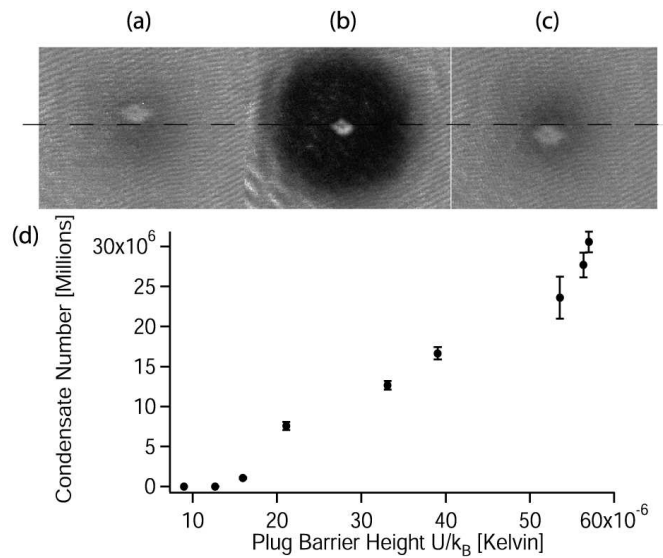}
\end{center}
\vspace{-6mm} \caption{Plugging the hole.  Catastrophic loss
results if the ``plug'' is not carefully aligned.  In b) it is
aligned precisely with the magnetic field zero (indicated by a
dashed line), resulting in a large number of atoms near the end of
the rf evaporation stage, while in a) and c) it is misaligned
along the y-direction by $+90 \mu$m and $-65 \mu$m, respectively,
resulting in very few atoms.  The plug laser power also
dramatically affects the number of condensate atoms achieved, as
shown in d), although large condensates were obtained even when
the plug is not completely capable of repelling atoms loaded from
the MOT, whose initial temperature was $T_i = 190 \mu$K.  The
field of view in each image is $1 \rm{mm} \times 1 \rm{mm}$.
\vspace{-4mm}} \label{fig:plug}
\end{figure}

We explored the crucial role played by the plug during
evaporation.  Figure 3 demonstrates how strong Majorana losses
are--when the plug beam was misaligned from the magnetic field
center by more than $50 \mu$m along any direction (Figs. 3a,3c),
tremendous losses ensued and very few atoms remained near the end
of the evaporation ramp.  However, when it was correctly aligned,
as in Fig. 3b, we obtained a huge increase in probe absorption, an
unmistakable signature that the hole had been successfully plugged
and that we could produce a BEC.

While the plug beam is clearly necessary, a natural question is
how high the potential barrier must be to prevent Majorana loss.
One estimate is that it should exceed the average energy of atoms
passing nearby.  This implies that $U_0\geq 3/2 k_B T_i$, where
$T_i$ is the initial temperature of atoms loaded into the
quadrupole trap.  The MIT group used a Stark shift of $350 \mu$K
for atoms at $200 \mu$K in the first plug trap. However, the
dependence on plug power was not determined in that work.  In Fig.
3d we plot the final atom number in the condensate versus plug
laser power.  At our maximum laser power we achieved condensates
of up to $30 \times 10^6$ atoms, close to the state-of-the-art
achieved in IP traps. Surprisingly, the peak Stark shift was only
$60 \mu$K, about 30 \% of $k_B T_i$, which we measured to be 190
$\mu$K. Thus the estimate above appears rather conservative.
Indeed, on occasion we have observed condensates of more than
$10^6$ atoms using a total power output from the laser of only 1
Watt, corresponding to $U_0/k_B T_i \simeq 0.1$. We interpret this
to mean that Majorana loss becomes critical only when the cloud
size becomes small.  Our observations suggest that by focusing
more tightly to a beam waist of $20 \mu$m or less, one may only
need a 0.5-1 watt plug laser.

Finally, we address a drawback of the hybrid magnetic and optical
trap--the exact potential energy surface depends on the precise
laser beam alignment {\em and} its profile in the wings,
parameters which are difficult to control.  We demonstrate that
the OPT can be used as a ``cooling stage'' to achieve quantum
degeneracy, followed by transfer of atoms (in a procedure similar
to \cite{stam98odt}) into a purely optical trap formed by a
single, focused infrared laser beam.  Here the atoms are localized
near the intensity {\em maximum}, where the potential energy
surface is well-defined. Our trap was formed by the output of a
1064 nm fiber laser which passed through an AOM and a single mode
optical fiber before being focused onto the atoms with a beam
waist of roughly $25 \mu$m.  In spite of a significant mode
mismatch between the OPT and the purely optical trap, we achieved
a transfer of up to 9 million atoms, or 50 \% of the condensate.

In conclusion, we have demonstrated a state of the art trap for
Bose-Einstein condensation.  We are currently working on spatial
filtering of the laser beam to create a ring trap, where new
superfluid phenomena can be expected, including persistent
currents \cite{muel98,rokh97ring}.  A unique feature of our
condensate is the non-uniform magnetization due to the
inhomogeneous quadrupole field, which may allow for the creation
of novel spin textures \cite{naka99,pu01}.

We thank W. Ketterle for useful discussions, A. Traverso, G.
Belenchia, P. Gabolde and B. Kaiser for experimental assistance,
and M. S. Chapman for a critical reading of the manuscript.  This
work was supported by the DoE and GT startup funds.

%\bibliography{References}% Produces the bibliography via BibTeX.

\end{document}